\documentstyle[12pt]{article}

\newcommand{\rf}[1]{(\ref{eq:#1})}
\newcommand{\op}{\partial\!\!\! \slash }

\begin{document}

\title{Geometric Phase in Quaternionic Quantum Mechanics}
\author{ M. D. Maia\\
Universidade de Bras\'{\i}lia, Instituto de F\'{\i}sica\\ 
70910-900 Bras\'{\i}lia, DF - Brazil\\
maia@fis.unb.br\\
V. B. Bezerra\\
Universidade Federal da Para\'{\i}ba, Departamento de Fisica\\
58051-970 Jo\~ao  Pessoa, Pb - Brazil \\
valdir@fisica.ufpb.br}

\maketitle

\begin{abstract}
Quaternion quantum mechanics  is   examined at the level of  unbroken  $SU(2)$
gauge symmetry. A general quaternionic phase  expression is  derived from
formal  properties of the quaternion algebra.

\end{abstract}

pacs {03.65.Bz, 03.65.Ca,11.15.Kc, 02.30.+g}

\section{Quaternion Quantum Mechanics}  

Quantum mechanics defined  over   general algebras have been conjectured  since 1934 \cite{Jordan}.  In 1936  Birkoff and von Neumann noted that  the propositional calculus implies  in a representation of  pure states of a quantum  system by rays on a Hilbert space defined over any associative division algebra  \cite{Birkoff}. This means that  quantum theory would be limited to the  real, complex and quaternion algebras.  
Standard textbooks  explain  the   complex formulation of  quantum
mechanics  by means of the double  slit experiment  and the complex phase
difference of the  wave functions. It is  possible to use a real quantum 
theory, but at the cost of introducing  a special   operator  $J$ satisfying   $J^{2}=1$ and $J^{T}= -J$, so that at  the end the  complex  structure emerges again.
The final argument for complex algebra as minimum requirement
appears with the spinor structure. In  fact,  the   spinor  representations of the  
rotation group requires the  existence of solutions  of   quadratic 
algebraic equations  related to  the  invariant operators, which  
are guaranteed  only  over a complex algebra\cite{Chevalley}. 

The development of quaternion quantum mechanics 
started with D. Finkelstein in 1959,  its relativistic and  particle  aspects   were studied  by  G. Emch  and  E. J.  Schremp \cite{Fink,Emch:1,Schremp}. A  comprehensive reference list  can be found in \cite{Adler}. 

In an attempt to interpret  quaternion quantum mechanics, C. N. Yang
suggested  that the  isospinor  symmetry  should be contained in the group of 
automorphisms of the  quaternion algebra \cite{Fink}. Indeed, supposing
 that the spin angular momentum $\vec{M}$  associated with  $SO(3)$ and the  isospin   $\vec{I}$  given by a representation of  $SU(2)$, are both present in a single state,
their  spinor representations are given by the Pauli matrices acting  separately   on the  spinor  space  ${\cal M}$ and the  isospinor space ${\cal I}$  respectively,   generated by two independent complex bases   $(1,i)$ and  $(1,j)$. The direct sum  ${\cal M}\oplus{\cal I}$
does not close as   an algebra, except if   third  imaginary unit  $k=ij$ is introduced,
producing a quaternion algebra. The automorphisms of this algebra carries the  spin-isospin combined symmetry.

According to this interpretation, quaternion quantum mechanics would be
effective at the energy level in which the spin and isospin symmetries remain
combined.  When  this combined symmetry breaks down,  the isospin angular momentum
may lead  to  an extra  spin  degree of freedom \cite{JR,Hooft,Vachaspati,Singleton,Emch:2}. 

The  existence and effectiveness of  quaternion quantum mechanics 
at higher energies must be experimentally  verified. In one of the  experiments  
proposed by   A. Peres,  a neutron interferometer with thin 
plates made of  materials with varying proportions of neutrons and 
protons  is  used,  where the phase difference in one or another case is  measured \cite{Peres}. In principle this  experiment  could be  adapted to a  variable 
beam intensity, so that a   phase difference between  the complex 
and the quaternion could be detected for different energy levels. 

Quaternions  keep a  one-to-one correspondence with space-time 
vectors. Therefore,  the  quaternion phase  can  also be set in a one-to-one
correspondence with a rotation subgroup of the Lorentz group. In this sense, the  geometric quaternion phase is  truly geometrical as compared  with the  geometric complex phase, defined on a projective space \cite{Berry}.
The integration of this  quaternionic  geometric phase,  along a closed loop in  space-time,  can be associated with the  space-time curvature,   suggesting a quantum gravitational  effect. 

Taking the  quaternion wave function $\Psi$  as  a 
solution of Schr\"odinger's equation defined  with an anti-Hermitian 
quaternionic Hamiltonian operator $H$, then the quaternionic dynamical 
phase $\omega$  can be   described by \cite{AA} 
\begin{equation}
\oint_{c}{ \omega}^{-1} {d \omega}\; = -\int_{C} <\Psi |H |\Psi>dt 
\label{eq:qphase}
\end{equation}
where  $<,>$ denotes the  quaternionic Hilbert product. In \cite{AA},
Adler and Anandan have proposed a solution of  this integral as   given 
by 
\begin{equation}
{ \omega}(t)\; = T \ exp\left( -\int_{0}^{t} < \Psi |H|\Psi>dv \right), 
\label{eq:qphase2}
\end{equation}
where $exp$ is  the usual complex  exponential function and  $T$  denotes a  constant quaternion  representing  a time ordering factor.

On the other  hand,  the  geometric phase  $\tilde{\omega} $is  determined 
by  
\begin{equation}
\oint_{c}{\tilde \omega}^{-1} {d \tilde \omega} = -\int_{C} < \Psi|
{\frac{d \Psi}{dt}}>dt, 
\label{eq:qphase3}
\end{equation}
Again,  according to\cite{AA} this  may be   integrated to  give the following result
\begin{equation}
{\tilde \omega}(t) = T \exp ({-\int_{0}^{t} 
< \Psi |{\frac{d \Psi}{dv}}>dv}). 
\label{eq:qphase4}
\end{equation}

The question  we address ourselves   concerns with the generality of this  solution  as represented  by   a   Volterra integral  combined with a  fixed 
ordering factor $T$.  We  shall see that  the left-hand  side of  \rf{qphase3}  has a  solution  expressed by  the imaginary  quaternionic  exponential  function.  
We  start by examining the  meaning and  uniqueness of the   quaternionic line integral in \rf{qphase3}.
Denoting a quaternion function of a quaternion variable 
$X= X_{\alpha}e^{\alpha}$   in a quaternion basis $e^{\alpha}$ by
$f(X)=U_{\alpha}e^{\alpha}$ where  $U_{\alpha}$ are  real  components, we may define  the left  and  right line  integrals respectively  by\footnote{Greek indices run from 0 to  3 and small Latin indices from 1 to 3.  The  quaternion multiplication table is
$e^{i}e^{j} =-\delta^{ij} +\sum\epsilon^{ijk}e^{k}$,  $e^{i}e^{0}  =e^{0}e^{i}=e^{i}$ and  $e^{0}e^{0}=e^{0}=1$. Quaternion conjugate is  denoted with overbar: $\bar{e}^{i} =-e^{i}$,   $\bar{e}^{0}=e^{0}$. The  quaternion norm is  $|X|^{2}=X\bar{X}$.}   
\[
\int_{c} f(X)dX  =
e^{\alpha}e^{\beta}\int_{c}U_{\alpha}dx_{\beta}, \;\;\;\;
\int_{c} dX f(X) = 
e^{\beta}e^{\alpha}\int_{c}U_{\alpha}dx_{\beta}.
\]
These integrals  are not necessarily  equal:
\[
\int_{c}f(X)dX -\int_{c} dX f(X)=  \epsilon_{ijk}e^{k}\int_{c} (U_{i}dx_{j}
-U_{j}dx_{i}).
\]
However,  for a  closed loop  $c$, the  above  difference vanishes as a consequence of Green's theorem in the plane $(i,j)$.
Therefore,  for  $ X=\tilde{\omega}$ and  $ f(X)=\tilde{\omega}^{-1}$, the phase expression  in the left-hand side of \rf{qphase3} is  uniquely  defined.

To  obtain a solution  of  \rf{qphase3}   that is more general than \rf{qphase4}    we need to  understand the importance of being a  division algebra.

\section{Harmonic Functions}

The  division algebra condition  $|AB|=|A||B|$ is   a  basic  requirement 
of the  standard mathematical analysis based on   limit operations of products.
 In the complex case, the limit operation  is independent of the direction in the Gauss plane, eventually leading to the Cauchy-Riemann  equations.
However,   when we attempt  to   extend  the same  concepts to the  
quaternion algebra, the generalized Cauchy-Riemann conditions become
so restrictive that only  a few trivial functions survive (see the appendix for  a  brief review)   \cite{Fueter,Ketchum,Ferraro,Nash,Khaled}. 
A less restrictive condition is   given by the  harmonic  functions \cite{Gursey}
\begin{equation}
\sum\delta^{ij}\frac{\partial^{2} U_{\alpha}}{\partial X_{i}\partial
X_{j}}+\frac{\partial^{2} U_{\alpha}}{\partial X_{0}^{2}}=\Box^{2} U_{\alpha}=0,\label{eq:box}
\end{equation}
Quaternion harmonicity  can be implemented by  the introduction  of the slash differential  operator  $\op =\sum  e^{\alpha}\partial_{\alpha}=\sum e^{\alpha}{\partial}/{\partial X_{\alpha}}$, such  that   $\Box^{2}=\op\bar{\op}$.  
Clearly this operator  may act  on the  right and on the left  of a quaternion
function $f(X)$, giving 
\begin{eqnarray}
\op f(X) & = & (\frac{ \partial U_{0}}{ \partial X_{0}} +
\sum \frac{\partial U_{i}}{\partial X_{0}}e^{i})\nonumber\\
        & +  & \sum[\frac{\partial  U_{0}}{\partial X_{i}}e^{i}
- \sum \frac{\partial U_{i}}{\partial X_{j}}(\delta^{ij}
-\epsilon^{ijk}e^{k})], \nonumber\\
f(x)\op & =  & (\frac{\partial U_{0}}{\partial X_{0}} +\sum \frac{\partial
U_{i}}{\partial X_{0}}e^{i}) \nonumber\\
       &  + & \sum[\frac{\partial  U_{0}}{\partial X_{i}}e_{i} 
-\sum \frac{\partial U_{i}}{\partial X_{j}}(\delta^{ij}+ 
\epsilon^{ijk}e^{k})]\nonumber
\end{eqnarray}
It is clear that  $\op f(X)\neq f(X)\op $, unless the condition
\begin{equation}
\frac{\partial U_{i}}{\partial X_{j}}= \frac{\partial U_{j}}{\partial X_{i}},
\end{equation}
holds. Therefore, three  classes of harmonic functions may be defined:\vspace{3mm}\\
a) The left  harmonic functions, characterized by $ \op f(X)= 0$
\begin{eqnarray*}
  \frac{\partial U_{0}}{\partial X_{0}} &=&
\sum_{i} \frac{\partial U_{i}}{\partial X_{i}},\\
   \frac{\partial U_{k}}{\partial X_{0}}  &+&\frac{\partial U_{0}}{\partial
X_{k}}
= \sum_{ij}\epsilon^{ijk}\frac{\partial U_{i}}{\partial X_{j}}.
\end{eqnarray*}
b) The right  harmonic functions, such that $f(X)\op\!=0\!$
\begin{eqnarray*}
&& \frac{\partial U_{0}}{\partial X_{0}}=
\sum_{i} \frac{\partial U_{i}}{\partial X_{i}},\\
&& \frac{\partial U_{k}}{\partial X_{0}} +\frac{\partial U_{0}}{\partial X_{k}}
\!=\!-\!\sum_{ij}\epsilon^{ijk}\frac{\partial U_{i}}{\partial X_{j}}.
\end{eqnarray*}
c) The   totally harmonic functions (or  simply H-functions),  characterized by 
$\op f(X)\!=\! 0\;\mbox{and}\; f(X)\op\! =\!0\! $:  
\begin{eqnarray}
\frac{\partial U_{0}}{\partial X_{0}}
&=&\! \sum_{i}\frac{\partial U_{i}}{\partial X_{i}},\nonumber\\ 
\frac{\partial U_{i}}{\partial X_{0}}
&= &\! -\!\frac{\partial U_{0}}{\partial X_{i}}\label{eq:H}\\
\frac{\partial U_{i}}{\partial X_{j}}
&=&   \frac{\partial U_{j}}{\partial X_{i}}.\nonumber
\end{eqnarray}
The functions  belonging to these three classes  satisfy  the harmonic  condition \rf{box}.

A non trivial  example of  H-function is   given by an
instanton   expressed in terms of  quaternions. The
connection  of  an anti  self dual $SU(2)$ gauge field is given by the  form \cite{Atiah}
\begin{equation}
\Gamma =\sum_{\alpha}A_{\alpha(X)}dx^{\alpha}, \label{eq:harmonic}
\end{equation}
where   $A_{0}=\sum U_{k}e^{k}$ and  $A_{k}=U_{0}e^{k} 
-\epsilon_{ijk}U_{i}e^{j} $ and  \[U_{0}=  \frac{\frac{1}{2} 
X_{0}}{1+|X|^{2}},\;\;\;U_{i}=\frac{-\frac{1}{2}X_{i}}{1+|X|^{2} },\]
are   the components of the  quaternion function  $f(X)=U_{\alpha}e^{\alpha}$.
We can   see that   $f(X)$  satisfy the  conditions  \rf{H}
in the  region of  space-time  defined by $\sum X_{i}^{2}=-2X_{0}$. 

The  above  example is  a particular case  of  a  wider class of  functions
with components 
\[
U_{\alpha} =g_{\alpha}(X)/(1 +|X|^{2}),
\]
where $g_{\alpha}(X)$ are real  functions. 

However,  there are some functions which are  clearly analytic, such as  a constant quaternion, which does not satisfy \rf{H}. Therefore, as it happens in the  cases of
real and complex functions,   an  analytic  quaternion  function should be
more generally  defined by a convergent positive power series.

\section{Power Series}

The power expansion of  a quaternionic function  requires the equivalent 
to Cauchy's  integral  theorems. 
Given  a quaternion function $f(X)$ defined on an orientable 3-dimensional
hypersurface  $S$ with unit normal vector $\eta$, again we may define two
hypersurface  integrals.  
\[
\int_{S} f(X) dS_{\eta},\; \mbox{and}\,\;\int_{S}  dS_{\eta} f(X),  
\]
where   $dS_{\eta}=\sum dS_{i}e^{i}$  denotes the quaternion hypersurface
element with components
\[
dS_{\alpha}=\epsilon_{\alpha\beta\gamma\delta}dX_{\alpha}dX_{\beta}dX_{\gamma}
\]
where $\epsilon_{\alpha\beta\gamma\delta}$ is the four-dimensional Levi-Civita symbol.
On the other hand,
denoting by  $dv=dX_{0}dX_{1}dX_{2}dX_{3}$  the 4-dimensional  
volume element in a region $\Omega$ bounded by $S$,
after integrating in one of the variables, we obtain the following result
\begin{eqnarray*}
& &\int_{\Omega} \op f(X)dv= 
\int_{\Omega} e^{\alpha}\partial_{\alpha} e^{\beta}U_{\beta} dv =\\
&&\int_{\Omega}[(\partial_{0}U_{0}\!-\!\sum_{i}\partial_{i}U_{i}) \!+\!
\sum_{i} (\partial_{0}U_{i}+\partial_{i}U_{0}) e^{i} \! +
\!\epsilon^{ijk}\partial_{i}U_{j}e^{k}] dv.
\end{eqnarray*}
Noting that
\[
\int \partial_{\alpha}U_{\beta}dv=\int U_{\beta}dS_{\alpha}
\]
it follows   that
\begin{eqnarray*}
\int_{\Omega} \op f(X)dv &=&  \int_{S}
 \left[( U_{0}dS_{0}-\sum\delta^{ij} U_{i}dS_{j})e^{0}\right.\\
 & +& \left.\sum(U_{i}dS_{0}
+U_{0}dS_{i} )e^{i} -\sum \epsilon^{ijk} U_{i}dS_{j}e^{k} \right] .
\end{eqnarray*}
An straightforward calculation shows that  this is  exactly the same expression
of the  surface integral 
\[
\int_{S}dS_{\eta}f(X)=\sum \int_{S} U_{\alpha} dS_{\beta} 
e^{\beta}e^{\alpha}
\]
Therefore,  we obtain the result
\begin{equation}
\int_{\Omega} \op f(X)dv=\int_{S} dS_{\eta} f(X)  \label{eq:RDIV}.
\end{equation}
Similarly,   for the left  surface integral we have
\begin{equation}
\int_{\Omega}f(X)\op dv=\int_{S} f(X)dS_{\eta}.\label{eq:LDIV}
\end{equation}
These  integrals  are defined for  any  quaternion functions 
whose components are  integrable and their   difference is  
\[
\sum\epsilon^{ijk}e^{k}\!\!\int_{S}(U_{i}dS_{j}\!-\! U_{j}dS_{i})\!=\!
\!-\!\sum\epsilon^{ijk}e^{k}\!\!
\int_{\Omega}(\frac{\partial U_{i}}{\partial X_{j}} + \frac{\partial
U_{j}}{\partial X_{i}})dv,
\]  
The right hand side  is zero so that   only one type of   surface  integral  need to be considered. The following theorem extends the  first Cauchy's Theorem to  quaternion  functions:

{\em If $f(X)$ satisfy  \rf{H} in the interior of  a region $\Omega$
bounded by a hypersurface $S$ then }
\begin{equation}
\int_{S} f(X) dS_{\eta} =\int_{S}dS_{\eta} f(X)=0. \label{eq:CAUCHY1}
\end{equation}
This property follows immediately from equations \rf{RDIV}, \rf{LDIV} and 
the conditions \rf{H}.
The second  Cauchy's theorem is  also true  for  H-functions: 

{\em If $f(X)$  satisfy the conditions \rf{H} in a  region  bounded by  a
simple closed 3-dimensional hypersurface $S$,  then for  a  point $P $ in 
$ S $,we have}
\begin{equation}
f(P)=\frac{1}{\pi^{2}}\int_{S} f(X)(X-P)^{-3}dS_{\eta}. \label{eq:CAUCHY2}
\end{equation}
The proof  is also a  trivial  generalization  of the  similar complex theorem:
The integrand  does not satisfy the   conditions \rf{H} in $\Omega$
as it is not  defined   at $P$ and consequently
the previous theorem does not apply. However the   point   $P$  may be isolated by a
sphere with  surface  $S_{0}$ with center at $P$ and radius  $\epsilon$ 
such that it  remains inside  $\Omega$.  
Applying   \rf{CAUCHY1}  to   the region bounded by $S$  and  $S_{0}$
we obtain
\[
\int_{S} f(X)(X-P)^{-3}dS_{\eta} +\int_{S_{0}}f(X)(X-P)^{-3}dS_{\eta} =0.
\]
Now, the  components  $U_{\alpha}$  may be assumed to be differentiable 
and regular so that   we may  calculate their Taylor expansions  around  $P$: 
\[
U_{\alpha}(X) =  U_{\alpha}(P) +\epsilon^{\beta}\frac{\partial
U_{\alpha}}{\partial x^{\beta}}\rfloor_{P} +\cdots .
\]
Replacing   this  in the   integral over  $S_{0}$ and taking the limit
$\epsilon\rightarrow 0$,  it follows that
\begin{equation}
f(P)\! =\!\left( \int_{S} f(X)(X\!-\! P)^{-3}dS_{\eta}\right) \left(
\int_{S_{0}} (X\!-\! P)^{-3}dS_{\eta}\right)^{-1}. \label{eq:fp}
\end{equation}
In order to calculate the integral over the sphere $S_{0}$  it is convenient to use four
dimensional spherical coordinates  
$(r,\theta, \phi,\gamma) $,  such that
$X_{0}=r sin\gamma $,  $X_{1}=r cos\gamma\, sin\theta\,cos\phi $,
$X_{2}=r cos\gamma\, sin\theta\,sin\phi $ and  $X_{3}=r cos\gamma\, cos\theta$
where  $\theta\in(0,\pi)$, $ \phi\in(0,2\pi)$, $ \gamma\in(-\pi/2 , \pi/2)$.
The volume element in spherical coordinates   is  $dv=J dr d\theta d\phi 
d\gamma$ where $J=-r^{3}cos^{2}\gamma\sin\theta$  is the Jacobian determinant. 

The  unit normal  to the spherical hypersurface
centered at $P$ and with radius $\epsilon$ can be 
written as  $\eta=(X-P)/\epsilon$  so that $(X-P)^{-3}
={\bar{\eta}^{3}}/{\epsilon^{3}}$  and
\[
-\int_{S_{0}} (X-P)^{-3}dS_{\eta}=\int_{S_{0}}
{\bar{\eta}^{2}}cos^{2}\gamma \sin\theta\, d\theta\, d\phi\,
d\gamma=\pi^{2}.
\] 
After replacing in \rf {fp}, we obtain  the proposed result.
Notice that  the  power  $(-3)$  in \rf{CAUCHY2} is not accidental
as it is  the right  power  required to  cancel  the Jacobian
determinant when  $\epsilon\rightarrow 0$.

 Now we  may   prove  the following general  result:

{\em Let $f(X)$ be such that it satisfies \rf{H} inside a region $\Omega$ 
bounded by a surface $S$.  Then for all $X$ inside  $\Omega$   there exists
coefficients  $a_{n}$ such that }
\begin{equation}
f(X) =\sum_{0}^{\infty}  a_{n}(X-Q)^{n}.  \label{eq:TH3}
\end{equation}
As in  the  similar complex theorem, consider   the largest sphere $S_{0}$  inside $\Omega$, centered at $Q$. The  integral  \rf{CAUCHY2}  for  a point  $P=X$ inside $\Omega$ gives
\[
f(X)\! =\! \frac{1}{\pi^{2}}\int_{S}\!
f(X')(X'\!-\! Q)^{-3}[1\!-\! (X'\!-\!Q)^{-1}(X\!-\!Q)]^{-3}dS'_{\eta}. 
\]
It is  a simple matter to see that  the particular function  $f(X)=(1-X)^{-3}$, with
$\vert X\vert<1$   can  be expanded as 
\begin{equation}
(1\!-\! X)^{-3}\!=\!\sum_{1}^{\infty}\frac{n(n+1)}{2}X^{n-1}\! =\!
\sum_{m=0}^{\infty}\frac{(m\!+\!1)(m\!+\!2)}{2}X^{m}.\label{eq:example}
\end{equation} 
Assuming that $|X-Q| < |X'-Q|$ and using  \rf{example}, the above integrand is equivalent to 
\begin{eqnarray*}
&&[1-(X'-Q)^{-1}(X-Q)]^{-3}=\\
&&\sum_{0}^{m=\infty}\frac{(m+1)(m+2)}{2}(X'-Q)^{-m}(X-Q)^{m}, 
\end{eqnarray*}
so that 
\begin{eqnarray*}
&&f(X)=\frac{1}{\pi^{2}}\sum_{m=0}^{\infty}\frac{(m+1)(m+2)}
{2}\times\nonumber\\
&& \int_{S_{0}}f(X')(X'-Q)^{-3-m} (X-Q)^{m} dS'_{\eta}. 
\end{eqnarray*} 
Since $\eta$ and  $(X-Q)$ are proportional,  the above expression may be written as 
\begin{eqnarray*}
&&f(X)=\frac{1}{\pi^{2}}\sum_{m=0}^{\infty}\frac{(m+1)(m+2)}{2}\times\\
&& \int_{S_{0}}f(X')(X'-Q)^{-3-m}dS'_{\eta}\, (X-Q)^{m}, 
\end{eqnarray*}
Defining  the coefficients 
\begin{equation}
a_{m}\!=\! \frac{1}{\pi^{2}}\frac{(m\!+\!1)(m\!+\! 2)}{2}
\int_{S_{0}}f(X')(X'\!-\! Q)^{-3-m} dS'_{\eta}, \label{eq:COE1}
\end{equation}
we obtain  expression   \rf{TH3}, showing  that  all  functions satisfying \rf{H} can  also be  expressed as  a convergent positive power  series. The converse is not generally true.

\section{Back to Phase}

Now  we  may define a  quaternion exponential function  in terms of  convergent power  series and in particular the   pure imaginary  quaternionic exponential to represent the quaternionic  phase.
Let us express the solution of \rf{qphase3} as the quaternionic ordered  exponential  function defined along a curve  $c$by
\[
\tilde\omega =
\mbox{qexp} (\int_{c}{\tilde{\omega}}^{-1} d\tilde{\omega}) 
\]
To find the   expression  of   $Pexp$,
consider the quaternion $X=X_{0}e^{0} +\sum X_{i}e^{i}$. With the last 
three components we may associate the 3-vector  $\vec{\xi}$, and a pure 
imaginary quaternion $\xi$ such that
$|\xi|^{2}=\sum X_{i}^{2} =\vec{\xi}\cdot\vec{\xi}$,
where the dot means the  Euclidean scalar product. 
The unit vector  $\vec{\Upsilon}=\vec{\xi}/\sqrt{\vec{\xi}\cdot\vec{\xi}}$,  corresponds to the pure imaginary quaternion $\Upsilon= \xi/|\xi|$, such that $\Upsilon^{2}=-1$, determined by
three  angles. 

We may now  draw the Gauss plane with $e^{0}$ in the real axis and  
$\Upsilon$ in a  direction orthogonal  to $e^{0}$. Then a quaternion  $X$  with modulus  $|X|$, making an angle $\gamma$ with $e^{0}$  may be expressed  as
\[
X=X_{0}e^{0} +X_{i}e^{i}=|X|\left( e^{0} cos\gamma +  \Upsilon sin \gamma   
\right) .
\]
Replacing $sin\gamma$ and  $cos\gamma$ by the respective power series expansions, after  rearranging the terms, we may define  the  quaternion  exponential $\mbox{qexp}(\Upsilon\gamma)$  by the  series  within  the parenthesis, so that
\[
\mbox{qexp} (\Upsilon\gamma)= \frac{X}{|X|}
\]
Therefore,  the   most general  integral of  \rf{qphase3} may be expressed as  the geometric  quaternion phase
\begin{equation}
\tilde\omega =\mbox{qexp}(\Upsilon\gamma) = e^{0} cos\gamma +\Upsilon sin\gamma. \label{eq:phase}
\end{equation}  
Notice that  in contrast to \rf{qphase4}  there is no fixed direction $T$ but  rather the  unit direction $\Upsilon$ which  varies  with the  quantum states. Contrarily to the complex  phase,  $\tilde\omega$  acts automorphically over the quaternion wave functions as 
\[
\Psi' =\mbox{qexp}(\Upsilon\gamma)^{-1}\Psi \,\mbox{qexp}(
\Upsilon\gamma)=\tilde{\omega}^{-1}\, \Psi\, \tilde{\omega},
\] 
Consequently,  the general quaternion phase  is in fact distinct from the complex 
phase both from the analytic  point of view  as well as  from its the geometric
interpretation.  In  the   particular case  where the  vector  $\vec\Upsilon$ is  fixed  we  obtain  a    solution  equivalent to \rf{qphase4}.

It appears that quaternion quantum mechanics  should be effective at the level of 
a combined  spin-isospin symmetry. The  quaternionic spinor  operator
transforms under the automorphism of the quaternion algebra, producing a distinct  behavior on the phase of the wave functions,  as compared with  the complex theory.

The neutron interferometry experiment proposed in
\cite{Peres}  can be modified to accommodate the  high-energy interpretation.
Accordingly, we suggest a variable  beam experiment over plates made of the 
same material. A higher energy beam will  show  a qualitative difference 
from the  lower  energy case,  evidencing the distinction between
 the quaternion and  complex phases.  
 
\newpage
\begin{center}
{\bf Appendix\\
{ \small Basic Quaternionic Analysis}}
\end{center}
Taking a  generic  quaternion function  $ f(X)=U_{\alpha}(X)e^{\alpha}$, and  denoting $\Delta f= [f(X+\Delta X) -f(X)]$, the left  and  right derivatives  of $f(X)$  are  defined respectively   by
\begin{eqnarray*}
f '(X) = \mbox{lim}_{\Delta X\rightarrow 0}\delta f(X) (\Delta X)^{-1},\\
'f(X) = \mbox{lim}_{\Delta X\rightarrow 0} (\Delta X)^{-1}\Delta f(X),
\end{eqnarray*}
where the  limits are  taken  with  $|\Delta X|\rightarrow 0$ 
along  the direction of the four-vector $\Delta X$  which depends on the
3-dimensional vector $\vec{\Delta X}$. 
Following the   same complex procedure, take the derivatives  along a  fixed direction  
$\Delta X =\Delta X_{\beta} e^{\beta}$ (no sum on $\beta$), indicated by 
the  index within parenthesis
\begin{eqnarray*}
f'(X)_{(\beta)}   =  \frac{\partial U_{0}}{\partial X_{\beta}}e^{0}
(e^{\beta})^{-1} + 
\sum_{i}\frac{\partial U_{i}}{\partial X_{\beta}}e^{i}(e^{\beta})^{-1},\\
'f(X)_{(\beta)}   =  \frac{\partial U_{0}}{\partial X_{\beta}}
(e^{\beta})^{-1}e^{0} + 
\sum_{i}\frac{\partial U_{i}}{\partial X_{\beta}}(e^{\beta})^{-1}e^{i}.
\end{eqnarray*}
Straightforward   calculation  shows that
\begin{eqnarray*}
f'(X)_{(0)} & = & \,  'f(X)_{(0)},\\
f'(X)_{(j)} & = & \,  'f(X)_{(j)}-2\sum_{i,k}\epsilon^{ijk}\frac{\partial
U_{i}}{\partial X_{j}} e^{k} 
\end{eqnarray*}
By imposing  that these derivatives  are selectively  equal,
four basic  classes of complex-like analytic  functions can be obtained:

\begin{center}

\begin{tabular}{||c|c|c|c||}\hline\hline
 right analytic   &  left analytic & left-right analytic  &   total analytic\cr\hline
&  &   &\cr
 $f'(X)_{(0)} \! = f'(X)_{(i)}$ & $ 'f(X)_{(0)}\!  = \, 'f(X)_{(i)}$  &   $f'(X)_{(\alpha)}=  \, 'f (X)_{(\beta)}$ &        $f'(X)_{(\alpha)}\!  =\! f'(X)_{(\beta)}$,
\cr
 $f'(X)_{(i)} \! =  f'(X)_{(j)}$  &  $'f(X)_{(i)}\!  =\,   'f(X)_{(j)}$  &                              &  $ 'f(X)_{(\alpha)}\!  =\! \, 'f(X)_{(\alpha)}$
 \cr
                                     &                        &                             &    $'f(X)_{(\alpha)}\!=\! f'(X)_{(\alpha)}$ 
     \cr & & &  \cr\hline
&  &   & \cr
$\frac{\partial U_{\alpha}}{\partial X_{\alpha}} = \frac{\partial U_{\beta}}{\partial X_{\beta}} $ &  $\frac{\partial U_{\alpha}}{\partial X_{\alpha}}= \frac{\partial U_{\beta}}{\partial X_{\beta}}  $ &     $\frac{\partial U_{\alpha}}{\partial X_{\alpha}}= \frac{\partial U_{\beta}}{\partial X_{\beta}}  $ & $\frac{\partial U_{\alpha}}{\partial X_{\alpha}} = \frac{\partial U_{\beta}}{\partial X_{\beta}}  $ 
\cr &   &   &\cr
$\frac{\partial U_{i}}{\partial X_{0}} = -\frac{\partial U_{0}}{\partial X_{i}}$       & 
$\frac{\partial U_{i}}{\partial X_{0}}= -\frac{\partial U_{0}}{\partial X_{i}}  $       &
$\frac{\partial U_{i}}{\partial X_{j}}= -\frac{\partial U_{j}}{\partial X_{i}}    $        &
$\frac{\partial U_{\alpha}}{\partial X_{\beta}} =  0\;\;  \alpha\neq \beta        $  
\cr &  &  & \cr
$\frac{\partial U_{i}}{\partial X_{j}} = \sum\epsilon^{ijk}\frac{\partial U_{k}}{\partial X_{0}}$  & 
 $\frac{\partial U_{i}}{\partial X_{j}}= -\sum\epsilon^{ijk}\frac{\partial U_{k}}{\partial X_{0}} $ &
 $\frac{\partial U_{i}}{\partial X_{0}}= -\frac{\partial U_{0}}{\partial X_{i}}    $                            & \cr
&  &   & \cr
\hline\hline
\end{tabular}

\end{center}

 As we have mentioned these  conditions are  too restrictive
for most  applications, including quantum mechanics.

\end{document}